\documentstyle[preprint,aps, epsfig,graphicx,epsf]{revtex}
\begin{document}
\title
{The ferroelectric Mott-Hubbard phase of organic $(TMTTF)_{2}X$ conductors.}
\author{P.~Monceau$^1$, F.~Ya. Nad$^{1,2}$ and S.~Brazovskii$^{3,4}$}
\address{$^1$Centre de Recherches sur les Tr\`es Basses Temp\'eratures,
laboratoire associ\'e\\
 \`a l'Universit\'e Joseph Fourier, CNRS, BP 166, 38042 Grenoble cedex 9, France\\
$^2$Institute of Radio-Engineering and Electronics, 103907 Moscow, Russia\\
$^3$Laboratoire de Physique Th\'eorique et des Mod\`eles Statistiques,\\
CNRS, B\^at.100, Universit\'e Paris-Sud, 91405 Orsay, cedex, France\\
$^4$ L.D. Landau Institute for Theoretical Physics, Moscow, Russia.
}
\maketitle
\begin{abstract}
We present experimental evidences for a ferro-electric transition in the
family of quasi-one-dimensional conductors $(TMTTF)_2X$. We interpret this
new transition in the frame of the combined Mott-Hubbard state taking into
account the double action of the spontaneous charge disproportionation on
the $TMTTF$ molecular stacks and of the $X$ anionic potentials.


\end{abstract}

\narrowtext
Low dimensional electronic systems serve as a workshop on problems of strong 
correlations. The richest opportunities have been opened by the family of charge
transfer salts $M_2X$ formed of stacks of organic molecules
tetramethyltetrathiofulvalene $(M=TMTTF)$ or tetramethyltetraselenafulvalene 
$(M=TMTSF)$ with anions $X=PF_{6},AsF_{6},SbF_{6},SCN$, etc as counterions.  The 
ions sit in loose cavities delimited by the methyl groups of the organic 
molecules. These materials  show almost all known electronic phases: a metal, a 
paramagnetic insulator (PI), spin/charge density waves (SDW/CDW), the spin-Peierls state and finally the superconducting state \cite{jerome-94}. 
In parallel, there is a set of several different structural types due
to anion ordering (AO) which are slight arrangements in $X$ chains 
\cite{anions}. Their transition temperatures $T_{a}\approx 100-200K$ are 
much higher than that corresponding to electronic transitions occurring in the 
range of $T_{c}\approx 1-10K$.

Very recently, in the already very rich phase diagram of these salts, a new
phase transition has been discovered, being revealed by a huge anomaly in
the low frequency dielectric constant $\epsilon ^{\prime }$ at $T_{0}=70K$ in 
$(TMTTF)_{2}PF_{6}$ and at $100K$ in $(TMTTF)_{2}AsF_{6}$ \cite{eps-99}. 
Very soon NMR studies have proved \cite{brown} that the
transitions at $T_{0}$ are related to charge disproportionation, also called
charge ordering (CO), dividing molecules within in two non equivalent species. 
Previous reports on $TMTTF$ salts with larger anions
such as $SbF_{6}$ have suggested the occurrence of a phase transition,
sometimes called `` structureless`` transition, at a temperature of the
scale of $T_{a}$ showing itself in changes of the resistance \cite
{lawersanne}  and the thermopower \cite{coulon}, in weak features in high 
frequency ($\sim GHz$) dielectric susceptibility \cite{javadi}. 
However, $X-$ rays investigations brought no observation of superlattice 
reflections \cite{lawersanne} while some variation of intensity of Bragg reflections could actually have been noticed \cite{q=0}. NMR allows to identify the very fact of the CO via nonequivalence of molecular sites 
\cite{brown,kanoda} but it cannot distinguish between the  $\vec{q}=0$ state and 
the known undulating ones (either along the stack like the $4k_F$ (Wigner) 
condensation in $(DCNQI)_2Ag$ \cite{kanoda} or between stacks like in 
$(TMTTF)_2SCN$) \cite{anions}).
 So, the identification of the nature of this `` structureless`` $\vec{q}=0$
transition was still unsolved. We show hereafter that $\epsilon ^{\prime }$
in $(TMTTF)_{2}SbF_{6}$, as in the case of $(TMTTF)_{2}PF_{6}$ and 
$(TMTTF)_{2}AsF_{6}$, exhibits a huge divergence at $154K$. Dielectric
experiments are providing thus an unique access to the anomalous collective
properties of this mysterious transition identifying the resulting state as a 
never expected ferroelectric (FE) one.

In the high temperature range $T>T_{a},T_{0}$ the counterions already have a
profound effect on the electronic properties: the anions dimerize the
intermolecular spacing of the TMTTF molecules along the stacks. As a result,
the conduction band is split into a filled lower band separa  ted from a
half-filled upper band by a dimerization gap, leading to Mott-Hubbard charge
localization. The effects brought by the AOs with $\vec{q}\neq
0$ have been previously discussed (\cite{anions,jetp} and refs. therein). The 
new effects of combined CO-AO with $\vec{q}=0$ are the main topic of the present 
work.
Empirically one can see a systematic difference between the usual 
$\vec{q}\neq 0$ and $\vec{q}=0$ transitions. The first ones have been always
observed for non centrosymmetric (NCS) anions \cite{Ag}; so, the orientational
ordering for which a short contact interaction between the anion and the
molecule is established was considered as the main mechanism, positional
displacements being only the consequence of it. In a contrary manner to 
$\vec{q}\neq 0$ AOs, the $\vec{q}=0$ ones are observed in systems with
centrosymmetric (CS) anions.

For such systems, we will present hereafter a possible universal mechanism
only related to a positional instability. The remarkable fit of the anomaly
in $\epsilon^{\prime} (T)$ for the three centrosymmetric salts , $X=PF_{6},AsF_{6}$ and 
$SbF_{6}$, to the Curie law suggests firmly the ferroelectric character of
these phase transitions. We account for this  FE state from
the combined action of charge disproportionation and ionic displacements.

The $(TMTTF)_{2}SbF_{6}$ were prepared by a standard electrochemical
procedure \cite{delhaes}. We present results on a sample with a regular
cross-section of $4\,10^{-5}cm^2$ and a length of 3 mm. Its room temperature
conductivity was about $10\,\Omega ^{-1}cm^{-1}$ close to the value previously
published \cite{lawersanne}. The complex conductance $G(T)$ was measured in
the frequency range $1kHz-10MHz$ using an impedance analyzer HP 4192A. The
amplitude of the ac signal was typically $10mV$, within the linear response.
With decreasing temperature, $G$ increases continuously (metallic type of
conductivity) to a maximum slightly above $T_{0}=154K$. Below $T_{0}$, $G$
sharply decreases with a thermally activated behavior with an activation
energy $\Delta =500K$. From room temperature to about $200K$ the magnitude of 
the $\epsilon ^{\prime }$ is relatively small (below our experimental resolution 
in this temperature range). With decreasing $T$
below  $200K$, we observed a sharp growth of the magnitude of $\epsilon
^{\prime }$ with a tendency to diverge near $T_{0}=154K$, reaching the huge
value of $\sim 10^{6}$ (at $100kHz$), and a very deep decrease of the
magnitude of $\epsilon ^{\prime }$ below $T_{0}$. Fig.1 shows the
temperature dependence of $1/\epsilon ^{\prime }$ in the 
$(TMTTF)_{2}X$ family with centrosymmetric anions $PF_{6},AsF_{6}$, \cite{eps-99} and 
$SbF_{6}$. The generic features are very similar for the three compounds: the
two branches above and below the appropriate $T_{0}$ are very close to be
linear, i.e. to follow the Curie law $\epsilon ^{\prime }$ $\sim /|T-T_{0}|$. 
The slopes of these branches ( i.e. the amplitudes of $A$) at $T<T_{0}$ 
are twice that at $T>T_{0}$, exactly as theoretically predicted in the
description of a second order phase transition.

One can get simply a general idea of the hidden ordering in considering the
joint effect of two sources to the dimerization, hence to the charge  gap 
$\Delta $. The {\em extrinsic} one is determined by the basic crystal
structure and {\em intrinsic} one is spontaneous being induced  by the electronic
sub-system itself. (This concept was intensively used in the study of
conducting polymers \cite{matv}). The gap $\Delta (W)$ appears as the
consequence of both contributions to the Umklapp scattering $W$. Without CO,
there is only the bond contribution $W_{b}$ to the amplitude $W$ from the
built-in bond alternation. The CO adds the on-site contribution $W_{s}$ from
the nonequivalence of sites. The charge gap is a function of the total amplitude 
$W=\sqrt{W_{s}^{2}+W_{b}^{2}}$ 
which results from the orthogonality of 
corresponding matrix elements of electronic scattering \cite{matv}. ( Since the 
gap is not a linear function of $W$, then the partial gaps do not add simply in quadrature as
the components of $W$). The energy change $F_{e}$ of the electronic system
due to both $W_{b}$ and $W_{s}$ depends only on the $\Delta $ and
consequently on the total $W$. But the energy of lattice distortions depends
only on the spontaneous site component $W_{s}$: 
$F_{l}=$ $1/2KW_{s}^{2}$.
Thus the total energy can be written in terms of the total $W$ as
$F_{tot}=F_{e}(W)+1/2KW^{2}-1/2KW_{b}^{2}$.
The ground state is determined by its minimum over $W$ under the restriction 
$W>W_{b}$. 
The spontaneous $W_{s}$ will appear if the two conditions are saticfied: 
a) the energy has a minimum at some value $W=W_{0}$ 
and b) $W_b$ is not too big to meet the restriction $W_{0}<W_{b}$. 
Since the value $W_{0}=W_{0}(T)$ increases with decreasing temperature, there will be a phase
transition at $W_{0}(T_0)=W_{b}$ provided that $W_{0}(0)>W_{b}$.

Now we are going to choose a microscopical description. Recall first a recent theoretical work \cite{seo} performed in the frame of a mean field (MF) approximation which has found that the CO can occur as the result of strong enough Coulomb interaction between electrons on the near-neighbor sites. This model has been developed at $T=0$ and it includes ultimately 
the spin ordering of AFM type which is characteristic for these systems at
low temperatures (about $10K$). The phase transitions we observe occur at
much higher temperatures ($100-200K$) and an appropriate approach is clearly
needed. Considering the case of real $3D$ materials, in our picture the CO is
stabilized by interaction with the $3D$ connected system of anions.
So in the following we will use the bosonization procedure which is suitable
for describing low energy and collective processes. 

Let consider a 1D electronic system with a mean occupation of $1/2$ electron
per site (that is per molecule) with sites along the stack labeled as $n$ at 
$x=na$.
The low frequency long wave effects are described by the Hamiltonian for the
phase $\varphi $ in the charge channel which is separated from the Hamiltonian for the spin channel. 
We shall normalize the phase according to conventions for CDWs and SDWs which are  
$2k_{F}=\pi /(2a)$ modulations $\sim \cos [\pi /2n+\varphi (x)+cnst]$. 
The $4k_{F}=\pi/a $ fluctuations of the charge density are $\rho
_{4k_{F}}\sim (-1)^{n}\cos [2\varphi (x)+cnst]$. For our crystals with two
molecules per unit cell the $4k_{F}$  is projected to $q_{x}=0$, then  $\rho
_{4k_{F}}$ describes the charge disproportionation within the unit cell.

The Hamiltonian for $\varphi $ (per site) can be written \cite{solyom} as: 
\begin{equation}
H_{0}+H_{u}\sim \frac{\hbar }{4\pi \gamma }
\left[ v_{\rho }(\partial_{x}\varphi )^{2}+v_{\rho }^{-1}(\partial _{t}\varphi )^{2}\right]
-Wcos(2\varphi )
\end{equation}
Without interactions $v_{\rho }=v_{F}$ and $\gamma =1$. Repulsive interactions 
reduce $\gamma <1$ which value carries all necessary information: either about 
on-site/inter-site interactions of the model \cite{seo} or about Coulomb 
interactions for the WC scenario like in \cite{kanoda}.
The Umklapp scattering amplitude $W$ of Luther and Emery or $g_{3}$ of Dzyaloshinskii
and Larkin \cite{solyom} is a feature of systems with one electron
per unit cell. Normally $W$ of the order of other interactions is not small;
hence a big gap is opened in the charge $\varphi -$ channel, so that only
gapless spin degrees of freedom are left for observations. But a specific
feature of $(TMTCF)_{2}X$ crystals is that $W$ is small and appear only as a
secondary effect of the anionic sublattice, opening an intriguing crossover
to "charge localization" \cite{jerome-94} (to PI in earlier language of 
\cite{jetp}).

The $4k_{F}$ susceptibility $\Pi _{4kf}\sim T^{4\gamma -2}$ is divergent
only for strong enough interactions when $\gamma <1/2$. For $\gamma >1/2$ the 
Umklapp term $\sim W$ is renormalized to zero and the
system is a $1D$ metal, nowadays called the Luttinger Liquid. At the
marginal line $\gamma =1/2$, known in terms of bare interactions as the
Luther - Emery line, $\Pi _{4kf}\sim \ln T^{-1}$ like for the Kohn anomaly
leading to the Peierls instability (then the Hamiltonian is reduced to spinless fermions 
which actually are $\pi $ solitons, with the gap $\Delta$). 
At arbitrary $\gamma <1/2$, $W$ is renormalized to a
finite value which we shall write as 
$W^{\ast }=\hbar a\omega _{t}^{2}/(8\pi \gamma v_{\rho })$,
$\omega _{t}\lesssim \Delta $. The frequency $\omega _{t}$
is a gap in the spectrum of linear phase excitations $\omega ^{2}=(v_{\rho
}k)^{2}+\omega _{t}^{2}$ and it enters, as a ''transverse'' frequency of the
optical response, the expression for the dielectric susceptibility 
\begin{equation}
\epsilon _{\Delta }=\gamma \frac{v_{\rho }}{v_{F}}
\frac{\omega _{p}^{2}}{\omega _{t}^{2}}\,,\;
\omega _{p}^{2}=\frac{8e^{2}v_{F}}{\hbar s}
\label{eps-Del}
\end{equation}
where $\omega _{p}$ is the plasma frequency of the parent metal and $s$ is
the area per stack. This contribution 
$\epsilon _{\Delta }\sim (\omega_{p}/\Delta )^{2}$ 
can be relatively large as $\sim 10^{3-4}$ and it
corresponds to the background upon which the anomaly at $T_{0}$ is
developed. But $\epsilon _{\Delta }$ is regular, showing only a dependence
on $\Delta $ which starts to increase below $T_{0}$ without signs of
decrease from above. These features would only add some upward kink in $\epsilon ^{\prime}$ below $T_{0}$ but cannot account for the observed anomaly.

The degeneracy between $\varphi =0$ and $\varphi =\pi $ corresponds to the displacement of the electronic system by just one lattice position $x$. Hence the $\pm \pi $ soliton just adds/removes one
electron in real space. The elementary excitations registered as charge
carriers are these solitons. Their activation energy $\Delta \sim \omega_{t}$ is 
determined from our results for  $G$ as e.g. $500K$ for $SbF6$. At finite
concentration of solitons as normal carriers $\sim \rho _{n}$ coming either
as thermal excitations or via incommensurability, there is an additional
contribution to the low frequency complex $\epsilon $ which we can write in
a Drude form 
$\epsilon _{n}\,=-\rho _{n}\omega _{p}^{2}/(\omega ^{2}+i\omega /\tau _{n})$ 
to be assigned  to the observed conductivity \cite{eps-99}. 

The scaling relations for the gap and the electronic energy read (in units of $E_F$)
$F_{e}\sim -\Delta^{2}\sim -W^\zeta$ 
with $\zeta=1/(1-\gamma)$
So the instability condition  is $\zeta<2$ that is $\gamma <1/2$. Indeed, at 
this condition the value $W$ can
appear spontaneously: the gain of the $F_{e}$ is higher than the loss $\sim
KW^{2}$ of energy due to spontaneous deformations being at the origin of the
potential $W$. The minimum is achieved at $W=W_{0}\sim K^{\beta}$ with 
$\beta^{-1}=2-\zeta =(1-2\gamma )/(1-\gamma)$.

For our particular system it is, by  principle, important to notice the two 
sources acting for the weak two fold commensurability, that is the two contributions to the
Umklapp interaction. The non dimerized system with $1/2$ electron per site
has a symmetry $x\rightarrow x+a$ and $x\rightarrow -x$ which corresponds to
phase transformations 
$\varphi \rightarrow \varphi +\pi /2$ and $\varphi \rightarrow -\varphi $. 
Thus the lowest order invariant contribution to the
Hamiltonian is $\sim W_{4}\cos 4\varphi$ which is usually negligibly small
as studies of CDW have shown. 
It is small as coming from Umklapp interaction of $8$ particles, half of them
staying far away from the Fermi energy; but also it is
renormalized as $\sim W_4^{4\zeta}$ so that superlow values $\gamma<1/8$ are 
required for its stabilization. 
For the site dimerization the symmetry $x\rightarrow x+a$ is broken while
the symmetries $x\rightarrow x+2a$ and $x\rightarrow -x$ are preserved. The
Hamiltonian, invariant under the corresponding transformations of $\varphi$: 
$\varphi \rightarrow \varphi +\pi $ and $\varphi \rightarrow -\varphi $ is 
$H_{W}^{s}=-W_{s}\cos 2\varphi $. The bond dimerization preserves the
symmetries $x\rightarrow x+2a$ and $x\rightarrow a-x$ 
(reflection with respect to the bond center). 
Hence the invariance is required with respect to $\varphi \rightarrow \varphi +\pi $ 
and $\varphi \rightarrow \pi /2-\varphi$ leading to $H_{W}^{b}=-W_{b}\sin 2\varphi $.

Altogether the nonlinear Hamiltonian becomes 
\begin{eqnarray}
H_{W} &=&-W_{s}\cos 2\varphi -W_{b}\sin 2\varphi =-W\cos (2\varphi -2\alpha
)\,;\;  \label{H_Ubs} \\
W &=& \sqrt{W_{b}^{2}+W_{s}^{2}}\,,\;\tan 2\alpha =W_{b}/W_{s}
\end{eqnarray}
For a given $W_{s}$ the ground state is doubly degenerate between $\varphi
=\alpha $ and $\varphi =\alpha +\pi $ which still allows for phase $\pi $
solitons. Moreover $W_{s}$ itself can change its sign between different
domains of anionic displacements. Then the electronic system must also
adjust its ground state from $\alpha $ to $-\alpha $ or to $\pi -\alpha $,
whichever is closer. Thus, at the domain boundary, a phase soliton of $ 
\delta \varphi =-2\alpha $ or $\pi -2\alpha $ will be placed carrying an non
integer charge $q=-2\alpha /\pi $ or $1-2\alpha /\pi $ per chain. The CO
modulation is 
$\rho _{co}\sim \cos (\pi x/a+2\alpha )=(-1)^{n}W_{s}/\sqrt{W_{s}^{2}+W_{b}^{2}}$.

The MF approach appears appropriate for anionic displacements:
critical fluctuations are suppressed by well pronounced 3D correlations, as
inferred from the perfect Curie like anomaly in $\epsilon ^{\prime }$ (Fig.1) 
as well as from X-ray studies on given AOs \cite{anions} . For electrons quantum 
fluctuations must be treated exactly
as it has been sketched above. The angle $\alpha$ is invariant and only the 
total gap value is subject to the renormalization from 
$W$ to $W^{\ast }(W)\sim \Delta^{2}$. 
Finally the total energy, as a function of the mean phase $\bar{\varphi}$ and 
$W_{s}$ at given $W_{b}$, is 
\[
F\sim -\Delta ^{2}\left( \frac{W_{s}}{W}\cos 2\bar{\varphi}+\frac{W_{b}}{W} 
\sin 2\bar{\varphi}\right) +\frac{K}{2}W_{s}^{2}-\frac{e}{\pi }E\bar{\varphi}
\]
which  includes contributions from electrons, lattice and electric field $E$. 
At $W_{b}>W_{cr}\sim K^{1/\beta }$  
this energy has only one minimum at $W_{s}=0$, $\bar{\varphi}=\pi /4$ 
(modulo $\pi $) but this point becomes unstable at $W_{b}=W_{cr}$ 
from which the anomaly in $\epsilon$
originates. At $W_{b}<W_{cr}$ this minimum is split in two ones which
appear, first closely, at 
$2\bar{\varphi}=2\bar{\varphi}_{\pm }=\pi /2\pm 2\delta \bar{\varphi}$, 
$W_{s}=W_{b}\tan 2\delta \bar{\varphi}_{0}$. 
The resulting ground state will be ferroelectric if the same 
$\bar{\varphi}_{\pm }$
is chosen for all stacks (our case), while the state is antiferroelectric (AFE) 
if the sign of $\bar{\varphi}_{\pm }$ alternates (the case of $(TMTTF)_2SCN$, see 
refs. in \cite{anions,jetp}. Above the CO instability the minimization yields 
\[
\varphi =\pm \frac{\pi }{4}-\frac{eE}{W_{b}-W^{cr}}\,,\;W_{s}=\frac{ 
E/2\pi }{W_{b}/W^{cr}-1}\,,\;\epsilon^{\prime} =\frac{4e^{2}/s}{W_{b}-W^{cr}}
\]
In writing $\epsilon^{\prime}$ we recall that $e\varphi /s\pi $ is the electronic
polarization which equals $\epsilon E$. We come back to the formula (\ref
{eps-Del}) but with a modified denominator: 
$\omega _{t}^{2}\rightarrow \omega _{t}^{2}-\omega _{cr}^{2}$.
Thus $\epsilon \rightarrow \infty $ at the transition point 
$W_{b}=W^{cr}$ 
which is the joint electron-ion instability. Importantly, the
value of $\Delta $ is related to the bare value of $\omega _{t}$ which stays
finite, hence the ''normal'' carriers (the $\pi -$ solitons) are gapful. 

Doing short, we have supposed above that instead of changing $T$ one
lets the parameters $K,W_{cr}$ or $W_{b}$ to be varied at $T=0$. Actually all
of them are functions of $T$ and the transition temperature $T_{0}$ is
determined by $W_{b}(T_{0})=W_{b}^{cr}(T_{0})$. Expanding around $T_{0}$ we
recover the experimental observation $\epsilon ^{-1}\sim T-T_{0}$ and we
need only to understand why the Curie law extends over a rather broad region of $\delta
T=T-T_{0}\approx 30K$. Clearly the expansion is valid when $\delta T$ is
small in comparing to the lowest energy scales which are the ionic cage energy $\sim 
10^{2}K$ \cite{anions} and $\Delta \sim 10^{2}K$. It seems that relative to these scales 
the expansion within $\sim 10^{1}K$ is well assured.  
Remarkably we arrive even at the right magnitude of the observed 
effect: $ \epsilon \sim 10^{4}T_{0}/(T-T_{0})$.

In conclusion, we have shown that $(TMTTF)_{2}X$ salts with centrosymmetric
anions $X=PF_{6}$, $AsF_{6}$ and $SbF_{6}$ undergo a phase transition at
which the dielectric constant shows a huge divergent peak. The ferroelectric
character of this phase transition is demonstrated by the Curie law behavior
of $\epsilon ^{\prime }(T)$ of the three compounds. These results also bring
the nature of the so-called structureless transition of the 
$(TMTTF)_{2}SbF_6$ salt with CS anions. 
The transition has been explained by the combined action of the 
uniform shift of anions with respect to the oppositely charged organic stacks
and of the with charge disproportionation appearing in each $TMTTF$ molecule
yielding thus a macroscopic ferroelectric polarization. 
Our approach provides a physically transparent phenomenological interpretation 
in terms of strongly fluctuating $4K_{F}$ density wave, i.e. a local Wigner 
crystal, subjected to a weak two fold commensurability potential. Up to now 
these FE transitions concern only the salts based on $TMTTF$ molecules for which 
$T_{0}$ is already in the regime of Mott Hubbard charge localization state.
The salts based on TMTSF are less $1D$ and are metallic in this temperature
range which preclude dielectric measurements. Nevertheless, the same type of the 
FE transition may exist, just being hidden or existing only in a fluctuating
regime. If such a fact would be confirmed by future structural or NMR
studies, then the whole analysis of intriguing abnormal properties in the
normal state should be revised, as it already needs to be made, see also \cite{brown} for the
PI phase of $TMTTF$ salts with centrosymmetric anions as
demonstrated by the present work.

We would like to thank C. Carcel and J. M. Fabre for providing us the
samples. Part of this work was supported by the Russian Fund of Basic
Research ( grant N 99-02-17364) and the twinning research program ( grant
N98-02-22061) between CRTBT-CNRS and IRE-RAS.

\begin{figure}
\includegraphics[angle=-90]{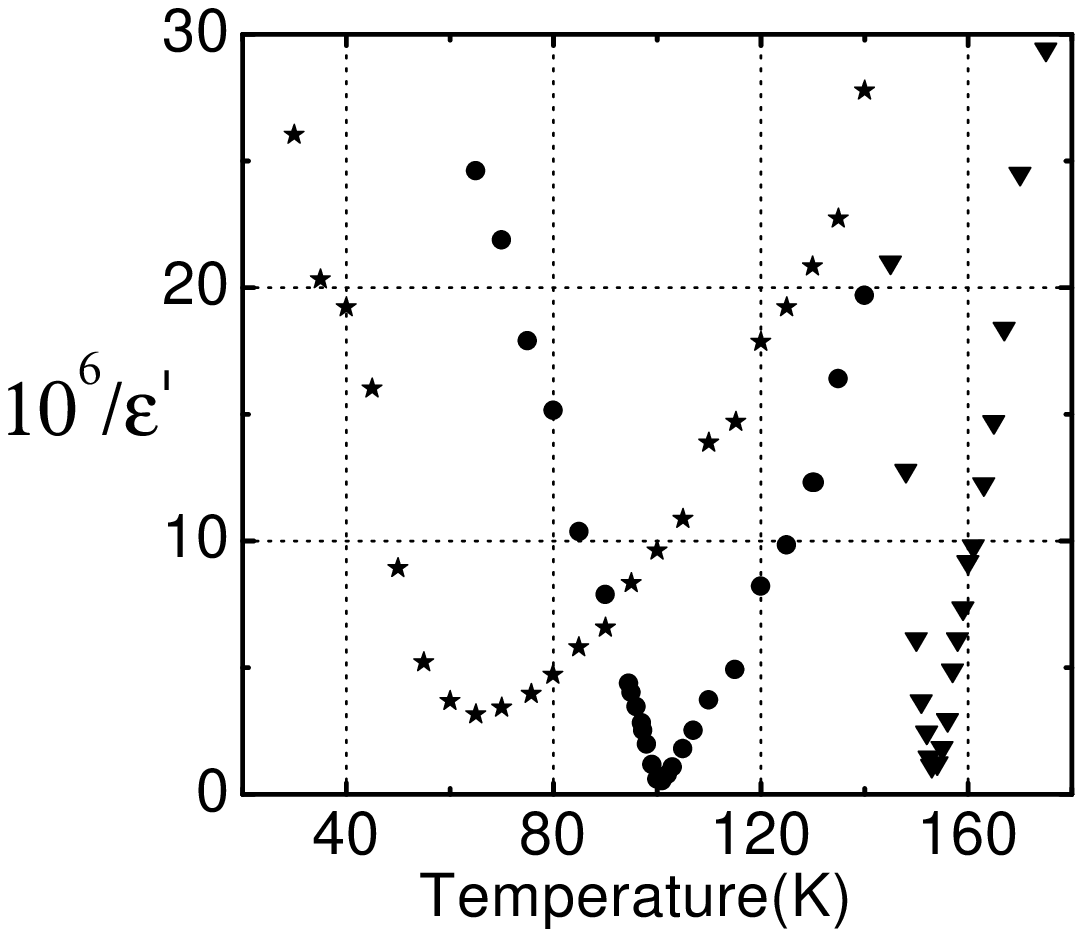}
\caption{
Temperature dependence of the inverse of the real
part of the dielectric permittivity $\epsilon ^{\prime}$ of $(TMTTF)_{2}X$ with 
$X= PF_{6}, AsF_{6}, SbF_{6}$ at the frequency of $100Hz$.
} 
\label{fig1}
\end{figure}


\begin{references}
\bibitem{Ag} In $(DCNQI)_2Ag$ \cite{kanoda}, the  
$\vec{q}\neq 0$ ordering is observed as a WC even for CS anyons.
\end{references}
\end{document}